\documentclass{aastex701}
\hypersetup{colorlinks=true, linkcolor=magenta, citecolor=magenta, filecolor=magenta, urlcolor=magenta}

\shorttitle{``Lensing is Low'' with DECADE}
\shortauthors{Queen, Fisher, \& Lange}

\begin{document}

\title{Revisiting the Mass and Scale Dependence of the ``Lensing is Low'' Effect with DECADE}

\author{Kimiko Queen}
\email{kq4453a@american.edu}
\affiliation{Department of Physics, American University, 4400 Massachusetts Avenue NW, Washington, DC 20016, USA}

\author[0009-0005-0530-0605]{Abigail V. Fisher}
\email{af9886a@american.edu}
\affiliation{Department of Physics, American University, 4400 Massachusetts Avenue NW, Washington, DC 20016, USA}

\author[0000-0002-2450-1366]{Johannes U. Lange}
\email{jlange@american.edu}
\affiliation{Department of Physics, American University, 4400 Massachusetts Avenue NW, Washington, DC 20016, USA}

\begin{abstract}
    Galaxy--galaxy lensing (GGL) is a critical probe of cosmology and galaxy formation. However, predicted GGL amplitudes tend to be $\sim 20-40\%$ higher than observations, primarily on small scales, a tension called the lensing-is-low problem. If not resolved through astrophysics or measurement systematics, this discrepancy may indicate a tension in our cosmological standard model. \cite{Lange2021_MNRAS_502_2074}, using foreground galaxies from the Baryon Oscillation Spectroscopic Survey (BOSS) and lensing measurements from the Sloan Digital Sky Survey (SDSS), presented evidence that this mismatch extends to large scales where astrophysical explanations are disfavored. We revisit this claim using lensing data from the Dark Energy Camera All Data Everywhere (DECADE) project. Compared to SDSS, the new DECADE measurements reduce the large-scale tension while leaving measurements on small scales mostly unchanged. Ultimately, these new results favor a scale dependence in the lensing-is-low problem.
\end{abstract}

\section{Introduction}

GGL is a measure of the mass distribution around galaxies. \cite{Leauthaud2017_MNRAS_467_3024} showed that observed lensing amplitudes tend to be lower than predictions when assuming Planck cosmic microwave background (CMB) cosmological parameters \citep{PlanckCollaboration2020_AA_641_6}, an effect called the ``lensing-is-low'' problem. Specifically, \cite{Leauthaud2017_MNRAS_467_3024} investigated the measured GGL signal around BOSS CMASS galaxies \citep{Reid2016_MNRAS_455_1553} and found that several independent models all overpredicted the signal by around $20 - 40\%$ at small physical separations, i.e., $\lesssim 10 \, \mathrm{Mpc}/h$. This and subsequent studies \citep{Lange2019_MNRAS_488_5771, Yuan2020_MNRAS_493_5551, Lange2021_MNRAS_502_2074, Amon2023_MNRAS_518_477, ChavesMontero2023_MNRAS_521_937, Heydenreich2025_arXiv_2508_5746, Campbell2026_arXiv_2606_3651} have investigated whether this discrepancy could arise from astrophysical effects, measurement systematics, or differences in assumed cosmology. Most studies on the lensing-is-low problem to date have found discrepancies on small scales but inconclusive results on large scales, often due to a lack of high precision measurements on those scales. An important exception is \cite{Lange2021_MNRAS_502_2074}, who analyzed the measured GGL amplitude around BOSS LOWZ galaxies using lensing measurements from SDSS \citep{Reyes2012_MNRAS_425_2610}. This study found a strong discrepancy on all scales, without a strong dependence in the ratio of observed to predicted lensing amplitude. Generally, astrophysical effects can modify the predictions on small but not large scales. If such a broader inconsistency is not explained by measurement systematics such as SDSS photometric redshift calibration, it would hint at tensions in our cosmological standard model, i.e., the so-called $S_8$-tension. Here, we revisit the analysis by \cite{Lange2021_MNRAS_502_2074} using updated lensing measurements from DECADE.

\section{Data \& Methods}

We utilize the same LOWZ samples as \citet{Lange2021_MNRAS_502_2074}, i.e., luminous red galaxies in the redshift range $0.2 \le z \le 0.35$. Galaxies are divided into low-, intermediate-, and high-stellar mass bins to investigate the mass dependence of the lensing-is-low problem. We utilize the same ``Wisconsin'', ``Granada'', and ``Legacy'' mass estimates for this division, leading to a total of nine different galaxy samples.

GGL measures the gravitational lensing distortions of background ``source'' galaxies around a set of foreground ``lens'' galaxies, i.e., LOWZ galaxies in our case. Ultimately, GGL is a measure of the physical mass distribution around lens galaxies as described by the excess surface density, $\Delta\Sigma$. We measure GGL using \textsc{dsigma} \citep{Lange2022_ascl_soft_4006}, replacing the SDSS source galaxies used in \citet{Lange2021_MNRAS_502_2074} with source galaxies from DECADE \citep{Anbajagane2025_OJAp_8_6158}\footnote{\url{https://dhayaaanbajagane.github.io/data_release/decade/}}. For DECADE source galaxies, we use the cosmology selection in the Northern Galactic Cap (NGC). Furthermore, we only select the two highest-redshift tomographic bins of DECADE (\texttt{MCAL\_SEL\_NOSHEAR} $\geq$ 3) which suppresses systematic effects such as boost factors and intrinsic alignments \citep{Lange2024_OJAp_7_57}. Individual source redshifts are not precisely known, so we use the effective redshift distributions $n(z_{\mathrm{s}})$ of sources \citep{Anbajagane2025_OJAp_8_6159} to calculate the effective critical surface density as a function of lens redshift $z_\mathrm{l}$,
\begin{equation}
    \Sigma_{\mathrm{crit, ls}}(z_\mathrm{l}) = \left[ \int_0^\infty n(z_{\mathrm{s}})\Sigma_\mathrm{crit}^{-1} (z_\mathrm{l}, z_{\mathrm{s}}) dz_{\mathrm{s}} \right]^{-1} \, .
\end{equation}
Note that $n(z_{\mathrm{s}})$ depends on the source tomographic bin. Next, we can construct an uncorrected estimate for the lens GGL amplitude from the weighted tangential ellipticities $e_\mathrm{t}$ of lens--source pairs at a given projected comoving separation $r_\mathrm{p}$ from the lens,
\begin{equation}
    \Delta\Sigma_\mathrm{L} = \frac{\sum_{\mathrm{ls}} w_{\mathrm{sys}} w_{\mathrm{ls}} e_{\mathrm{t}} \Sigma_{\mathrm{crit, ls}}(z_\mathrm{l})}{\sum_{\mathrm{ls}} w_{\mathrm{sys}} w_{\mathrm{ls}}} \, .
\end{equation}
In the above equation, $w_{\mathrm{sys}}$ is a systematic lens weight to correct for selection biases and $w_{\mathrm{ls}}$ is the lens--source weight designed to suppress shape noise. We obtain the final GGL estimate by subtracting the signal around random points $\Delta\Sigma_{\mathrm{R}}$, which suppresses additive systematics and cosmic variance \citep{Singh2017_MNRAS_471_3827}, and correcting for the average tangential shear response $R_\mathrm{t}$ and multiplicative bias $m$ in DECADE,
\begin{equation}
    \Delta\Sigma = \frac{1}{(1+\bar{m})\bar{R}_t}(\Delta \Sigma_{\mathrm{L}}-\Delta\Sigma_{\mathrm{R}}) \, .
\end{equation}
We use jackknife resampling to estimate the measurement covariance.

\section{Results \& Discussion}

\begin{figure}
    \centering\includegraphics{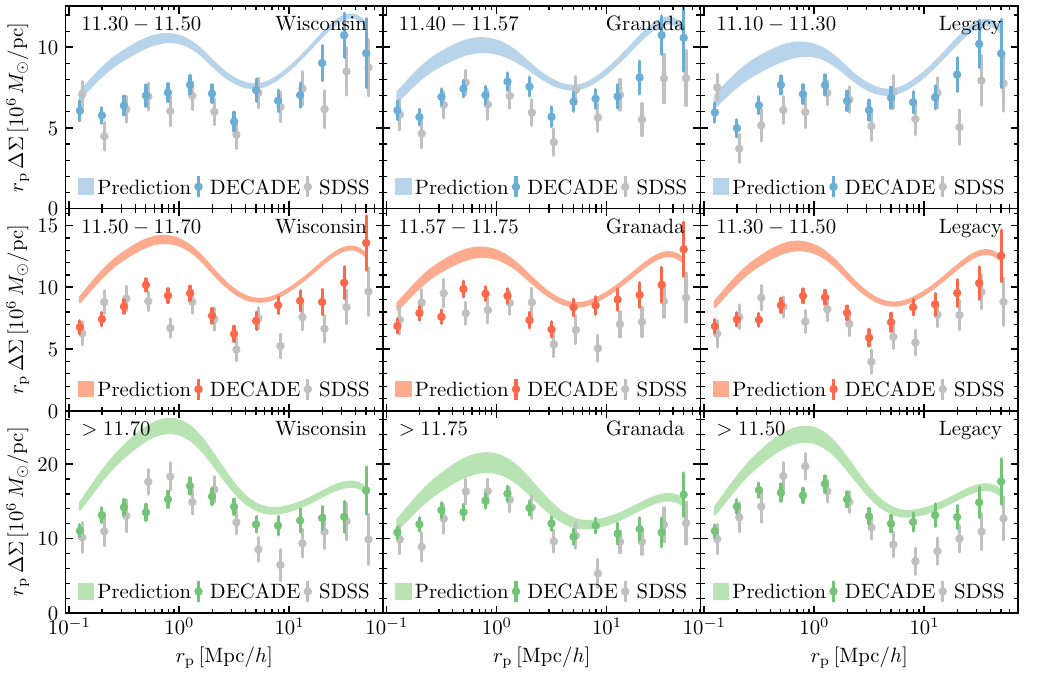}
    \caption{GGL measurements and predictions for the three stellar mass estimates. Colored bands show the $68\%$ confidence lensing predictions, and colored and gray points show the new DECADE and original SDSS measurements, respectively. Numbers in the top-left of each panel indicate the stellar mass bin in units of $\mathrm{log}_{10} M_\star / M_{\odot}$.}
    \label{fig}
\end{figure}

Fig.~\ref{fig} presents the new DECADE measurements together with the original SDSS amplitudes and model predictions from \citet{Lange2021_MNRAS_502_2074}. Compared to SDSS, the DECADE measurements show a similar discrepancy on small scales but approach the predictions more closely on larger scales. To assess a possible scale dependence, we model the ratio of observed to predicted lensing, $r = \Delta \Sigma_{\mathrm{obs}}/\Delta \Sigma_{\mathrm{pred}}$, assuming $r = b + m \,\mathrm{log}_{10} \, r_\mathrm{p}$. We use generalized least squares, taking into account uncertainties from measurements and predictions. When excluding scales below $r_\mathrm{p} < 0.25 \mathrm{Mpc} / h$ which are strongly affected by boost factors and the stellar mass of the lens galaxy itself, we infer positive slopes of up to $m \sim 0.14$ for all nine samples with significances ranging from $1.3\sigma$ to $3.8 \sigma$.

A scale dependence in the lensing-is-low problem would suggest that the discrepancy is at least in part due to astrophysical factors. On small scales, baryonic feedback and galaxy formation physics may contribute. Astrophysical processes can redistribute baryons, reducing halo densities on small scales, an effect not included in our predictions. Similarly, galaxy--halo connection models may not capture the full complexity of galaxy formation physics and lead to a further overprediction. Contrary, a discrepancy on large scales likely indicates a tension in our cosmological model, i.e., an $S_8$-tension. While our new DECADE measurements reduce the tension on large scales, the observations still tend to fall below predictions. More work is needed to determine whether the remaining difference is statistically significant in terms of cosmological structure formation predictions.

\software{\textsc{Astropy} \citep{AstropyCollaboration2022_ApJ_935_167}, \textsc{dsigma} \citep{Lange2022_ascl_soft_4006}, \textsc{matplotlib} \citep{Hunter2007_CSE_9_90}, \textsc{NumPy} \citep{VanDerWalt2011_CSE_13_22}}

\bibliography{bibliography}
\bibliographystyle{aasjournalv7}

\end{document}